\documentclass{article}
\usepackage{spconf,amsmath,graphicx,subcaption,cite}
\usepackage[utf8]{inputenc}
\usepackage[showseconds=false,showzone=false]{datetime2}

\def\f{{\mathbf f}}
\def\g{{\mathbf g}}
\def\s{{\mathbf s}}

\title{IMPROVEMENTS TO EMBEDDING-MATCHING 
ACOUSTIC-TO-WORD ASR \\
USING MULTIPLE-HYPOTHESIS PRONUNCIATION-BASED EMBEDDINGS
}

\twoauthors
 {Hao Yen\sthanks{Work done as an intern at Apple}}
	{Georgia Institute of Technology\\
	School of Electrical and Computer Engineering\\
	Atlanta, GA, USA\\
	rick.yen@gatech.edu}
 {Woojay Jeon}
	{Apple\\
	One Apple Parkway\\
	Cupertino, CA, USA\\
	woojay@apple.com}

\begin{document}

\ninept

\maketitle

\begin{abstract}
In embedding-matching acoustic-to-word (A2W) ASR, every word in the vocabulary is represented by a fixed-dimension embedding vector that can be added or removed independently of the rest of the system. The approach is potentially an elegant solution for the dynamic out-of-vocabulary (OOV) words problem, where speaker- and context-dependent named entities like contact names must be incorporated into the ASR on-the-fly for every speech utterance at testing time. Challenges still remain, however, in improving the overall accuracy of embedding-matching A2W. In this paper, we contribute two methods that improve the accuracy of embedding-matching A2W. First, we propose internally producing multiple embeddings, instead of a single embedding, at each instance in time, which allows the A2W model to propose a richer set of hypotheses over multiple time segments in the audio. Second, we propose using word pronunciation embeddings rather than word orthography embeddings to reduce ambiguities introduced by words that have more than one sound. We show that the above ideas give significant accuracy improvement, with the same training data and nearly identical model size, in scenarios where dynamic OOV words play a crucial role. On a dataset of queries to a speech-based digital assistant that include many user-dependent contact names, we observe up to 18\% decrease in word error rate using the proposed improvements.
\end{abstract}

\section{Introduction}
\label{sec:intro}

One of the main challenges in state-of-the-art neural-network-based automatic speech recognition (ASR) is the ability to handle user-dependent out-of-vocabulary (OOV) words, especially named entities like contact names, song titles, and business names \cite{huang-2020, zhao-2019, bruguier-2019, li-2017}.

An ``embedding-matching'' acoustic-to-word (A2W) ASR \cite{settle-2019, collobert-2020} is different from other A2W ASR systems \cite{audhkhasi-2017, soltau-2017, audhkhasi-2018} in that the final pre-softmax transformation matrix of the connectionist temporal classification (CTC) or encoder-decoder network is populated by an external text encoder.
Each column of the matrix is an acoustic embedding vector \cite{jeon-2020, he-2017} of a word in the vocabulary that represents how the word sounds.
The idea is that if such a network is trained over enough data while keeping the final matrix fixed, the output of the preceding layer will always be the acoustic embedding of an audio segment hypothesized by the network to contain a word.
At testing time, the pre-softmax transformation matrix can be arbitrarily extended with other word embeddings for any utterance.
This extensibility provides us a simple, scalable method for handling OOV words, particularly \emph{dynamic} OOV words such as user-dependent named entities that differ for every utterance and must be added to the ASR's vocabulary \emph{on the fly}.

However, not much work has been reported so far on the embedding-matching approach in the literature.
There seems to be a significant accuracy gap between embedding matching A2W ASR and other more mature state-of-the art ASR systems, such as FST-based hybrid DNN-HMM systems \cite{mohri-2008, zhen-huang-2020} that do on-the-fly composition with slot lexicon FSTs \cite{novak-2012, paulik-2016} to handle user-dependent OOV words. 
Hence, many challenges lie ahead in making embedding matching A2W competitively accurate.

In this study, we make two contributions to the endeavor of improving embedding-matching A2W.
First, we propose modifying the internal structure of the neural network such that \emph{multiple} embeddings are produced at every point in time, allowing the system to hypothesize on more diverse time segments.
Second, we propose the use of word-pronunciation (as opposed to word-orthography) embeddings to better handle words that have more than one sound. 

For all our experiments, we use acoustic neighbor embeddings \cite{jeon-2020}, which were shown to match the accuracy of FST-based DNN-HMM hybrid ASR in isolated word matching experiments.
Note, however, that the improvements proposed in this paper are not dependent on the specific type of embedding used, and may be equally applicable to other types of acoustic word embeddings (e.g. \cite{he-2017}).
Experimental results show that our proposed improvements bring about an 18\% decrease in word error rate on a test set of utterances spoken to a speech-based digital assistant containing numerous user-dependent contact names.

\section{Baseline System}
\label{sec:baseline}

\subsection{Conception}

We begin with a description of our baseline embedding-matching A2W system that is similar to a previously-reported ``frozen prediction layer'' system \cite{settle-2019} where a matrix of embedding vectors obtained from an external text encoder is used to compute the pre-softmax scores of a word CTC model.

\begin{figure}[t]
  \centering
  \includegraphics[width=2.5in]{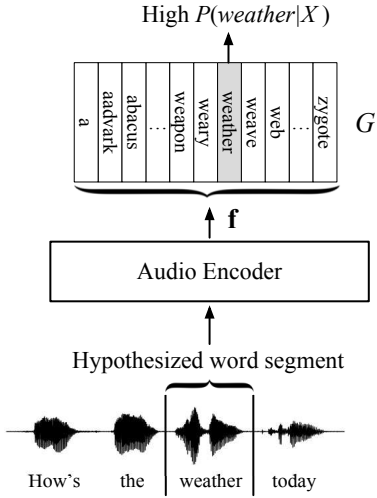}
  \caption{An illustration of the intuition behind embedding-matching A2W. A ``word segmenter'' scans the audio to identify segments that possibly correspond to individual whole words. Each segment is transformed into an audio embedding vector that captures the segment's phonetic content. The vector is compared against all text embedding vectors in matrix $G$, and (in the above example) best matches the vector for ``weather'', hence resulting in a high posterior score for ``weather.'' The columns of $G$ can be changed to any vocabulary we desire.}
  \label{fig:weather}
\end{figure}

First, we show in Figure \ref{fig:weather} how embedding-matching A2W can be interpreted. The system can be understood as a continuous-speech extension of the embedding-matching isolated word recognition experiments done in previous work \cite{jeon-2020}. We imagine a ``word segmenter'' model scanning a speech utterance to identify possible word segments in the audio. For each hypothesized segment, an audio encoder (usually denoted by $f(\cdot)$ \cite{jeon-2020, he-2017}) transforms the audio segment into an embedding vector $\mathbf{f}$ that represents the segment's phonetic content. Separately, a text encoder (usually denoted by $g(\cdot)$) transforms a dictionary of words into a set of embedding vectors. The audio and text encoder are trained in tandem \cite{jeon-2020, he-2017} such that matching audio and text will map to the same embedding vector.\footnote{Note that the encoders must also be trained such that distances between embeddings reflect phonetic confusability as in \cite{jeon-2020}. A few easy examples can show us that random projections, for example, would not lead to proper recognition.} Hence, if the hypothesized segment contains the audio for ``weather'', its $\mathbf{f}$ vector will most closely match the column in $G$ that represents ``weather''. Each word posterior is modeled as the similarity between $\f$ and the word's embedding vector, so $\hat{P}(weather|X)$ gets a high score at this instance in time. A decoder would choose the high-scoring words to output a recognition result. In effect, the ASR is based on nearest-neighbor searches between the $\f$ vectors produced at different instances in time and the words in $G$.

When interpreted this way, the system is similar to embedding-matching whole-word segmental ASR \cite{shi-2021}, with one key difference: Instead of generating all embeddings for a range of different segments every time, the neural network makes ``educated guesses'' about prospective word segments and exposes only those guesses.

It is also apparent that $G$ can be arbitrarily changed to almost any vocabulary we wish.
For instance, user-dependent contact names can simply be added as new columns to a general-vocabulary $G$.
This is an extremely simple way of extending the ASR's vocabulary, and also scalable because brute-force vector nearest-neighbor search is highly parallelizable and efficient in modern hardware \cite{garcia-2010}.

\begin{figure}[]
  \centering
  \includegraphics[width=2.2in]{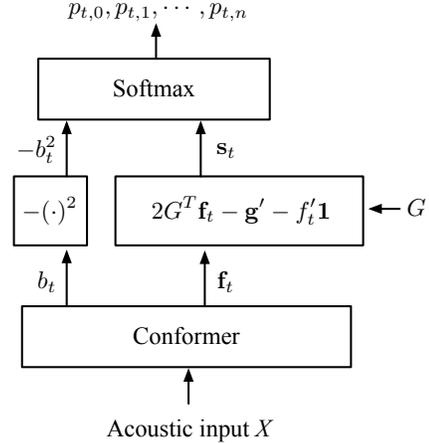}
  \caption{Baseline embedding-matching A2W system. The columns of the $G$ matrix contains the embedding vectors for all the words used during training. At every time $t$, the similarity score between the acoustic embedding $\f_t$ ``proposed'' by the conformer and every column in $G$ is outputted in $\s_t$, which is concatenated with the ``blank'' score $-b_t^2$ before the Softmax operation. The conformer model is trained by applying a CTC loss to the word posteriors, assuming a vocabulary of size $n$ ($p_{t,0}$ is the score for the ``blank'' word). The first dimension of the conformer's output is used as $b_t$, and the rest are used as $\f_t$ (an additional affine output layer ensures the correct number of dimensions is obtained).}
  \label{fig:baseline}
\end{figure}
 
In the actual system shown in Figure \ref{fig:baseline}, no such ``word segmenter'' explicitly exists, nor is an $f(\cdot)$ audio segment encoder even needed during training. Only a $g(\cdot)$ text encoder is used to populate the $G$ matrix, and the underlying network (a conformer \cite{gulati-2020} in this work) is trained via a CTC criterion. During training, the conformer is forced to output embeddings that match the appropriate columns in $G$, effectively learning to play both the roles of word segmenter and audio encoder as in Figure \ref{fig:weather}.

\subsection{Use of Euclidean distance and separate ``blank''}

In the baseline system in Figure \ref{fig:baseline}, at every time $t$ the conformer outputs a scalar $b_t$ and an acoustic embedding vector $\f_t$, which result in posterior scores for a word vocabulary $\{w_0, \cdots, w_n\}$, where $w_0$ denotes the conventional ``blank'' label \cite{graves-2006}:
\begin{equation}
    p_{t,i} = \hat{P}(w_i | X, t) \ \ \ (i=0, \cdots, n).
\end{equation}
We use acoustic neighbor embeddings \cite{jeon-2020} -- which were shown to match the accuracy of FST-based DNN-HMM ASR in isolated word matching -- for our $G$ matrix. 
Since the embeddings are trained based on the Euclidean distance rather than cosine distance, we use the negative square of the $\mathcal{L}_2$ distance, rather than the inner product in previous studies \cite{settle-2019, he-2017}, between the audio embedding $\f_t$ and each word embedding $\mathbf{g}_i$ to represent their phonetic similarity.
The similarity is maximized if and only if there is an exact match between the audio embedding and the text embedding, i.e., $\f_t = \mathbf{g}_i$.
\footnote{Note that, to achieve the same effect when using inner product, the embeddings must be constrained such that $||\f_t||=||\mathbf{g}_i||=c$ for some constant $c$. However, it is not clear whether the previous studies \cite{settle-2019, collobert-2020} did this.}

The pre-softmax output $\mathbf{s}_t$ stores the similarity scores at time $t$:
\begin{equation}
    \textbf{s}_t=s(\f_t, G)=2 G^T \f_t - \g' - f_t'\textbf{1},
\end{equation}
where $s(\cdot, \cdot)$ is the negative $\mathcal{L}_2^2$ distance between a vector and every column of a matrix, $\g'$ is a constant vector where every $i$’th element is $\g_i^T \g_i$, $f_t'=\f_t^T\f_t$, and $\mathbf{1}$ is a vector of 1s.

Previous studies \cite{settle-2019} defined an arbitrary embedding for the ``blank'' CTC symbol. In our study, we avoid this by directly using one of the outputs $b_t$ of the underlying neural network as the ``blank'' score. Since each score in $\s_t$ is $\le 0$ and essentially proportional to the square of each element of $\f_t$, we impose a similar restriction and ``scaling'' on $b_t$ by taking its negative square.

\section{Proposed Improvements}
\label{sec:proposed}

\begin{figure}[t]
  \centering
  \includegraphics[width=1.8in]{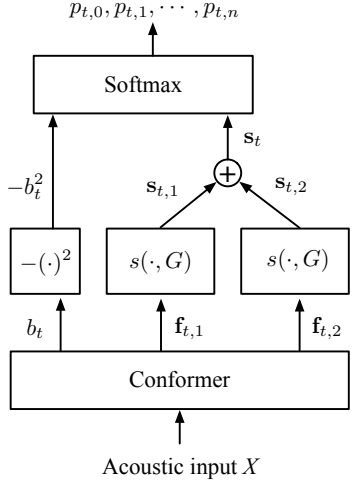}
  \caption{Proposed multiple-embedding-matching A2W system. The underlying network (in this case, a conformer) produces two embeddings instead of just one. Similarity scores are computed for each of the embeddings, then summed to simulate a ``logical OR'' operation that allows the model to simultaneously give high scores to two outputs at the Softmax layer.}
  \label{fig:proposed2}
\end{figure}

\begin{figure}[th]
    \centering
    
    \begin{subfigure}{0.5\textwidth}
    \includegraphics[]{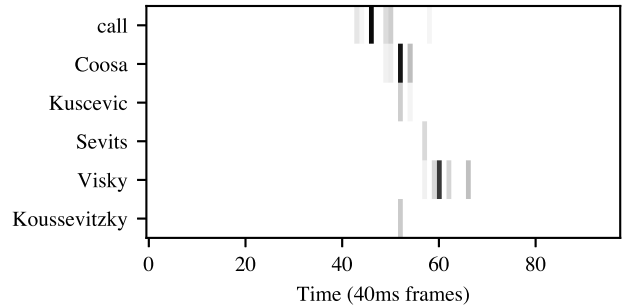}
    \subcaption{Scores from baseline single-hypothesis system in Figure  \ref{fig:baseline}}
    \label{fig:logposta}
    \end{subfigure}
    
    \begin{subfigure}{0.5\textwidth}      
    \includegraphics[]{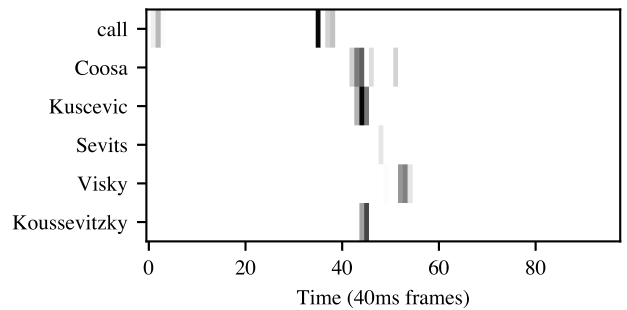}
    \subcaption{Scores from proposed multiple-hypothesis system in Figure  \ref{fig:proposed2}}
    \label{fig:logpostb}
    \end{subfigure}
    
    \caption{Log posteriors (dark is higher; for clarity, values below -20 are not shown) for the utterance ``Call Koussevitzky'' when using the baseline single-hypothesis system (Figure \ref{fig:baseline}) and the proposed multiple-hypothesis system (Figure \ref{fig:proposed2}). Even though both CTC models have nearly identical size and were trained on the same data, the latter produces more distinct scores for diverse segments in the audio, such as for the words ``Kuscevic'' and ``Koussevitzky''.}
    \label{fig:logpost}
\end{figure}

\subsection{Multiple embeddings}
\label{subsec:multiple}

A crucial limitation to embedding-matching A2W is apparent in Figure  \ref{fig:weather}, which is that the system can only hypothesize on one segment of the audio at any given point in time. For example, it is impossible for the system to simultaneously give high posteriors for both ``weather'' and ``wet.''
The reason is because ``wet'' has a shorter audio segment and different pronunciation compared to ``weather'', and therefore its audio embedding will be different, but the audio encoder can only produce one embedding at a time.

To address this limitation, we propose a method of producing \emph{multiple} embeddings at every point in time, and combining the resulting posteriors such that two words with different lengths and pronunciations may still simultaneously receive high scores.
This system is shown in Figure  \ref{fig:proposed2}. 
The final projection layer of the conformer is widened (the sizes of all other components remain the same) to produce two embeddings $\f_{t,1}$ and $\f_{t,2}$ at every time $t$. 
The scores against $G$ are individually computed, and then summed, simulating a ``logical OR'' operation where both $\f_{t,1}$'s matches against $G$ and  $\f_{t,2}$'s matches against $G$ can be outputted by the system.
One may also interpret this as a simplified mixture density network \cite{bishop-1994}.

An example is shown in Figure \ref{fig:logpost}, for the input audio ``call Koussevitzky.''\footnote{We randomly chose the name of this musician as an example.}
In the single-hypothesis system in Figure  \ref{fig:logposta}, we can see that the A2W has essentially limited itself to one fixed sequence of three audio segments corresponding to ``call'', ``Coosa'', and ``Visky''. Other words like ``Kuscevic,'' which overlaps with both ``Coosa'' and ``Visky'' only have weak scores. The faint score for ``Kuscevic'' is likely just a by-product from ``Kuscevic'' having a similar starting sound (and therefore a somewhat-close embedding) as ``Coosa.''

In the multiple-hypothesis system (3 embeddings) in Figure  \ref{fig:logpostb}, we see a more pronounced scores for a diverse set of segments. ``Kuscevic'' has a strong score, likely because the full length of its audio segment was actually hypothesized by the model. The same reasoning can be applied for the correct word ``Koussevitzky.''

Even though both the single-hypothesis and multiple-hypothesis networks have almost the same number of parameters (see Table  \ref{tab:comb1}) and are trained on the same data, we can get an improvement in versatility by internally restructuring the network in this manner.

\subsection {Word-pronunciation embeddings}

In a previous study \cite{jeon-2020}, it was shown that word pronunciation embeddings are generally more accurate than word orthography embeddings in isolated name recognition tasks. The former is obtained from grapheme (or character) sequences like [r, e, a, d], whereas the latter is obtained from phoneme sequences like [/r/, /iy/, /d/]. Because some words, like ``read'' or ``live'', can have more than one pronunciation, representing them with single embedding vectors can result in inaccurate matching. By using separate embeddings for their individual pronunciations, we can avoid such ambiguities.

Hence, in this study we also propose using pronunciation (or ``pron'') embeddings, where the model produces posterior scores for a vocabulary of word pronunciations instead of orthographies.

The model can be trained the same way as the baseline, but instead of providing a reference transcription of words, we provide a reference transcription of \emph{word prons}. Such a transcription can be obtained by force-aligning the audio to the reference transcriptions using a conventional hybrid DNN-HMM ASR with a pronunciation lexicon and finding the best pronunciation for each word.

When testing, the output pron posteriors $\hat{P}(r_0|X,t), \cdots,$ \\
$\hat{P}(r_m|X,t)$, assuming a pron vocabulary $\{r_0, \cdots, r_m \}$, is converted to word posteriors for a word vocabulary of size $n$ by setting
\begin{equation}\label{eq:word_post}
    \hat{P}(w_i|X,t) = \max_{r\in R_i} \hat{P}(r | X,t),
\end{equation}
where $R_i$ is the set of prons that word $w_i$ can have, e.g. if $w_1$=``live'', then $R_1$ = \{ ``/l/ /ih/ /v/'', ``/l/ /ay/ /v/'' \}. The ``blank'' posterior is used as-is, i.e., $\hat{P}(w_0|X,t)=\hat{P}(r_0|X,t)$.

In the case of homophones, like ``stair'' and ``stare,'' the posterior for ``/s/ /t/ /eh/ /r/'' becomes the posteriors for both words, and we leave it to the language model to choose the correct word.


\section{Experiment}
\label{sec:exp}

\begin{table}[th]
    \centering
    \begin{tabular}{ l | c| c | c }
     System & Size(M) & WER(\%) & NEER(\%) \\
    \hline
     Word-orth-1 (baseline) & 20.05 & 17.9 & 26.8 \\
     \hline
     Word-orth-2            & 20.06 & 15.5 & 22.8 \\
     Word-orth-3            & 20.07 & 15.4 & 22.9 \\
     \hline
     Word-pron-1            & 20.05 & 16.6 & 24.0 \\
     Word-pron-2            & 20.06 & 15.5 & 23.1 \\
     Word-pron-3            & 20.07 & 14.6 & 21.5 \\
    \end{tabular}
    \caption[]{Results using 1,000 hours of training data. The model size (millions of trainable parameters), word error rate (WER) and named entity error rate (NEER) are shown for the baseline and proposed systems. The baseline uses word-orthography embeddings while the proposed systems use word-pron embeddings. The number in the system name indicates the number of embeddings in Sec. \ref{subsec:multiple}. The same training data was used for all cases, and model sizes were near-identical, with a slight increase due to a larger projection matrix used at the end of the conformer when outputting multiple embeddings.} 
    \label{tab:comb1}
\end{table}

\begin{table}[th]
    \vspace{-.15in}
    \centering
    \begin{tabular}{ l | c | c | c | c }
     System & Hours & Size(M) & WER(\%) & NEER(\%) \\
    \hline
     Word-pron-3 & 3k  & 30.04 & 9.8 & 15.8 \\
     Word-pron-3 & 10k & 31.24 & 7.9 & 13.0 \\
    \end{tabular}
    \caption[]{Results using larger training data and models.} 
    \label{tab:comb2}
\end{table}

An external n-gram language model that uses a generic symbol
\emph{\texttt{\$CONTACT}} to represent contact names is used for all our evaluations. This symbol occurs in common contexts such as \emph{``call \texttt{\$CONTACT}''} or \emph{``send a text message to \texttt{\$CONTACT} I'm on my way,''} etc.
For every utterance, the list of contacts pertaining to the speaker is transformed to a set of embedding vectors and appended to $G$. A fairly generic prefix decoder (see, for example, \cite{hannun-2014}) ingests the final word posteriors (typically, only a few dozen with the top scores is enough). When the decoder sees any ``appended'' word, it treats it as \emph{\texttt{\$CONTACT}} for LM scoring purposes.

How to properly combine CTC model output scores with an external LM in a principled manner is still an open problem. In our study, we applied a known heuristic \cite{sak-2015} of simply dividing the ``blank'' posterior by a constant, and used the rest of the scores as-is. A usual language model scale factor was used to combine the word posterior in Equation  \eqref{eq:word_post} with a language model score.

All data was gathered from utterances spoken by diverse anonymized speakers to a digital assistant.
The evaluation data had 43k utterances, of which 14k contained at least one contact name.
In addition to word error rate (WER), the ``named entity WER (NEER)'' is measured on only the contact names (if any) in the test utterances.
For each contact in the reference, the corresponding word(s) in the hypothesis was located via minimum edit distance mapping, and the NEER was computed using only those words.

For the 1k-hour experiment in Table \ref{tab:comb1}, we trained conformer models \cite{gulati-2020} with 20M trainable parameters \footnote{4 heads, 16 layers, kernel size 31, and 232 dimensions} using a batch size of 256, and a frame subsampling factor of 2.
The input acoustic features were 80 mel filterbank outputs per frame, with an input frame rate of 10ms. 
All acoustic neighbor embeddings \cite{jeon-2020} $\f$ and $\g_i$ (in $G$) had 40 dimensions. The training vocabulary size was 13,266 for the word-orth case, and 12,860 for the word-pron case, obtained by gathering words (or prons) in the training data that occurred at least 10 times.
When testing, the vocabulary was replaced by a much larger vocabulary of 811,319 for the word-orth case and 811,485 for the word-pron case.
These were ``static'' vocabularies to which ``dynamic'' vocabularies, i.e., user-dependent contact names, were added for every utterance during testing.
The median number of contact names per user was 1,242.

In Table \ref{tab:comb2}, we increased the training data to 3k hours and 10k hours and tested on only word-pron embedding models generating 3 internal embeddings.
Conformer models had around 30M trainable parameters \footnote{8 heads, 18 layers, kernel size 31, and 272 dimensions}, and we used a batch size of 512, and a frame subsampling factor of 4.
The training vocabulary sizes were 32,193 for 3k hours and 62,231 for 11k hours.
The testing vocabulary size was the same as the 1k-hour experiments.

In Table  \ref{tab:comb1} and \ref{tab:comb2}, the accuracy improved significantly when we used multiple embeddings, with identical model sizes (i.e., number of trainable parameters) and training data. There was further marginal improvement when we switched from orthography embeddings to pronunciation embeddings. 

\section{Conclusion and Future Work}
\label{sec:conc}
In this work, we have proposed generating multiple embeddings in an embedding-matching A2W system, as well as using pronunciation-based embeddings, to make significant accuracy improvements to embedding-matching A2W.
More rigorous mathematical analysis can be done in the future to show how multiple embeddings are learned by the proposed system.

The embedding matching approach has yet to match the accuracy of state-of-the-art ASR systems that support dynamic OOV words. \footnote{For example, with a hybrid ASR using roughly similar data, we observed WER of around 5.5\% using 5k hours of training data.}
However, the ability to easily and arbitrarily modify the vocabulary of embedding-matching A2W in a scalable manner is attractive, and we believe it is worth exploring this direction further. We expect to report further novel  contributions and accuracy improvements to the method in the future.

\newcommand\blfootnote[1]{%
  \begingroup
  \renewcommand\thefootnote{}\footnote{#1}%
  \addtocounter{footnote}{-1}%
  \endgroup
}

\blfootnote{ACKNOWLEDGEMENT: Thanks to Zak Aldeneh, Siamak Dadaneh, Dhivya Piraviperumal, and Russ Webb for helpful comments.}


\bibliographystyle{IEEEbib}
\bibliography{refs}

\end{document}